\documentstyle[12pt,aaspp4]{article}
\begin{document}
\input psfig

\title{Localized Helioseismic Constraints on Solar Structure}
\author{John N. Bahcall}
\affil{Institute for Advanced Study, Princeton, New Jersey 08540}
\author{Sarbani Basu}
\affil{Theoretical Astrophysics Center,
Danish National Research Foundation,  \\
Institute for Physics and Astronomy, Aarhus University, DK 8000 Aarhus C,
Denmark}
\author{Pawan Kumar}
\affil{Institute for Advanced Study, Princeton, New Jersey 08540}

\begin{abstract}
Localized differences between the real sun and standard solar models
are shown to be small.  The sound speeds
of the real and the standard model suns typically 
differ by less than $0.3$\%  for regions of radial width $\simeq 0.1 R\odot$ 
in the solar core.   
\end{abstract}


\section{Introduction}

Since the first discrepancy was reported almost 30 years ago between 
observations 
of solar neutrinos (Davis, Harmer, \& Hoffman 1968) and the predicted fluxes of solar 
neutrinos
(Bahcall, Bahcall, \& Shaviv 1968),  
many authors have proposed {\it ad hoc} 
changes
in the input data to solar models or have 
hypothesized evolutionary scenarios in which the 
real sun differs from computed standard solar models.  Upon detailed 
examination,
the suggested modifications of the solar 
models have failed because 
of 
specific conflicts with other astronomical data or with laboratory 
measurements.
Nevertheless, the pressure to examine all allowable 
modifications of solar 
models has intensified as new solar neutrinos experiments, 
combined with more 
precise 
solar model predictions and the development of 
attractive particle physics 
explanations of the solar neutrino discrepancies,  
have made it clear that 
solar neutrinos might be revealing new physics beyond what is
described by the minimal 
standard electroweak theory.

In the last few years, helioseismic measurements have become 
increasingly precise  (see, e.g., Hill et al. 1996; 
Kosovichev et al. 1997)
and now include accurate measurements of the low degree
p-mode frequencies that extend  
well into the core of the sun ($r \leq 0.3 R\odot$) where nuclear 
fusion occurs and solar neutrinos are produced. Several groups have shown 
recently that the standard solar model predicts results for the sound 
speeds that are in remarkably good agreement with those measured by 
helioseismological techniques: the typical 
deviations between the models and 
the measurements are less than or of the 
order 0.2\% in the solar core and 
0.1\% rms throughout nearly all of the sun (see, Bahcall et al. 1997;
also  
Guenther \& Demarque  1997; 
Antia \& Chitre 1997; Kosovichev et al. 1997; Basu et al. 1996b;  
Gough et al. 1996).  

These previous studies show that the standard solar model represents
well the overall structure of the sun.
The sound-speed differences between the Sun and standard solar models
inferred from helioseismology are convolved with weighting kernels
that average the available information over appreciable volumes of the
sun (see, Figure~1 in the following section).  

In this letter, we set general constraints on 
possible localized deviations 
from standard solar models by making use of an accurate data set of 
helioseismic frequencies. We rule out localized   
differences
between sound speeds in the sun and in standard solar models that are
large enough to affect significantly  predictions of solar neutrino
fluxes.
For specificity, we consider Gaussian perturbations, but our numerical
results are expected to be also valid for oscillatory 
perturbations for which the wavelength is comparable to or larger than
the typical width of the averaging kernels (i. e., perturbation
wavelengths $\gtrsim 0.1 R_\odot$).

Helioseismic inversions yield values for 
the sound speed in the sun,  which are essentially proportional to 
$(T/\mu)^{1/2}$ (where $T$ is the temperature 
and $\mu$ the mean molecular 
weight). In the absence of unexpected fine-tuned 
cancellations between perturbations 
to $T$ and $\mu$, 
even tiny fractional errors in the models values of $T$ or $\mu$ would
produce measurable discrepancies in the precisely determined
helioseismic sound speeds.
Localized  modifications of the temperature 
which produce changes  exceeding a few tenths of a percent in the 
sound-speed profiles of standard solar models 
are ruled out in the regions tested  by existing 
helioseismic measurements. By contrast, the internal temperature 
in the standard solar model must be 
changed by typically 5\% to 10\% in order to affect in an important
way  solar 
neutrino calculations\footnote{As the referee points out, 
there is a mathematical possibility
that sound velocities in a non-standard solar model could
approximately equal the sound velocities in the standard model to an
accuracy of better than $0.2$\% , but nevertheless the temperature
differences are substantially larger.  
We do not investigate such scenarios in this paper
since no physical basis for so precise
 a fine tuning has ever been suggested.}
We concentrate here on assumed departures 
from the standard solar model that are confined to shells of 
characteristic 
widths of order $0.1 R_{\odot}$ since that is the typical width in 
which the 
solar neutrinos are produced. However, we also consider perturbations
in shells of width somewhat larger and somewhat smaller than 
$0.1 R_{\odot}$ and show that the 
observational upper limit to the amplitude of a sound 
speed perturbation, obtained from helioseismology, 
is inversely proportional 
to the thickness of the shell.

We present the results of our calculations in Section~2 and 
Figures~2 and 3 and then summarize and discuss the applications 
of our results in Section~3.

\section{Calculations}

The observed solar p-mode frequencies can be inverted to determine
localized averages of the relative sound-speed differences between 
the Sun and a specified solar model. 
The averages are taken over the {\it averaging kernels} which are
linear combinations of the functions that relate the 
sound-speed differences
between the Sun and a model to the differences of the frequencies.
The averaging kernels are constructed so that they are as narrow
as allowed by the data, and their finite radial widths define
 the resolution 
of the inversion. The narrower the averaging kernel, the closer is the
inferred  sound-speed difference to the 
true sound-speed difference at that
point. Details on how the p-modes are inverted and the
averaging kernels formed are given in Basu et al. (1996a).
The sidelobes of the kernels are small.

The averaging kernels we use for this work are those obtained by 
inverting 
solar oscillation frequencies obtained by the LOWL instrument 
(Tomczyk et al. 1995) in its first year of operations. 
Figure~1 shows a representative sample of the averaging kernels.
All p-modes in the LOWL data set with angular degree 
between $l = 0$ and 99 and
frequency in the range 1 to 3.5 mHz were included in the construction of
these kernels; the fractional errors in the frequency measurements for 
the LOWL data are of order a few times 10$^{-5}$.
Reliable kernels cannot yet be constructed in the subsurface area from
$0.95 R_\odot$ to $1.0 R_\odot$, but techniques have been found that
minimize the effects of the surface layers on the inferred interior
properties (see Basu et al. 1996a).  

\begin{figure}[t]
\centerline{\psfig{figure=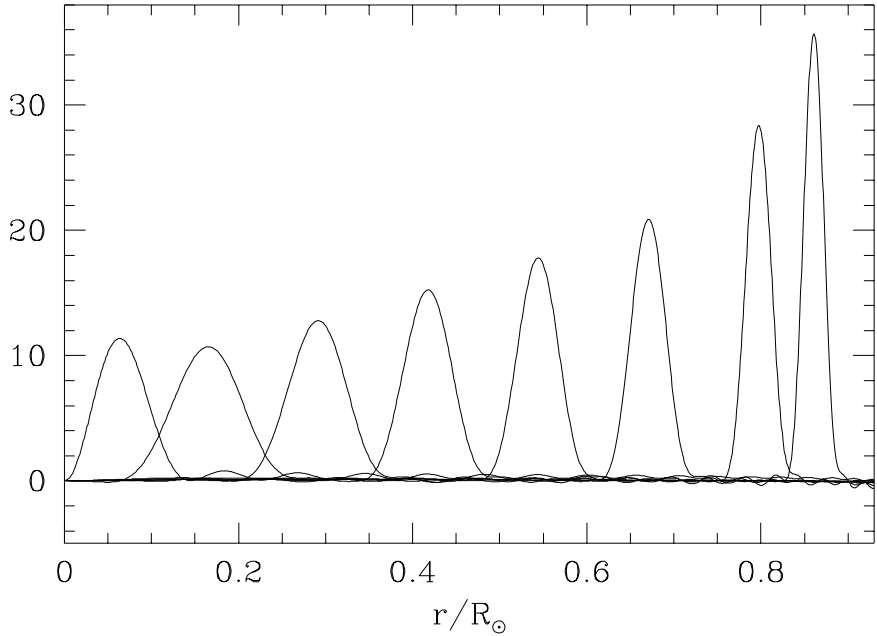,width=4.5in}}
\figcaption[bahcallfig1.ps]{A sample of the normalized averaging kernels for 
the relative sound-speed difference obtained using 
the first year's  data from the LOWL instrument (Tomczyk et al. 1995).
These correspond to the inversion results in  Basu et al. (1996b.)}. 
\end{figure}

The measurement errors lead to an error of about $0.05$\% in the
inverted sound-speed difference in the solar core and about $0.01$\%
elsewhere. 
The LOWL data set was used  to infer the difference in 
the sound-speed profile
of the Sun and that of a standard solar model (Basu et al. 1996b) and
has also been used to determine the sound-speed differences between 
the Sun and the Bahcall \& Pinsonneault solar model used to predict 
solar neutrino fluxes (see Bahcall et al. 1997).

By direct comparison, we find that the differences in 
the sound speeds computed from  the best standard solar model of 
Bahcall \& Pinsonneault (1995) and the solar Model S presented by  
Christensen-Dalsgaard et al. (1996) 
decrease from $0.3$\% at $R = 0.05 R_{\odot}$ to $0.1$\% at 
 $R = 0.15 R_{\odot}$.  Outside this innermost region of the sun, 
the differences in computed sound speeds between the two models are
always less than $0.1$\% and are typically 
less than a few hundredths of a per cent.
We conclude that these solar models constructed with
independent numerical codes and with different implementations of the
input physics nevertheless lead to almost the same sound speeds
Nevertheless, it will be important to determine in future work the
specific reasons for the discrepancies  between the two
standard models, especially 
 for $R \leq 0.15 R_{\odot}$  where the model-to-model variations
 are comparable to
the differences between the helioseismological values for the sound
speeds and the model values.

We have introduced artificial perturbations in the sound speeds at 
different solar radii and have evaluated the sensitivity of the 
helioseismic data to these departures from the standard solar model.  
Any change to the sound-speed profile will, of course, be  accompanied
by other changes to the structure that are required 
to satisfy the constraint of
hydrostatic equilibrium.  However, the inversion procedure is such that
we can invert for the  sound-speed perturbations only.

Figure 2 shows the results of a series of 7 perturbations introduced at  
different locations in the standard solar model, with the peaks of 
the perturbations appearing at 0.1, 0.2,  0.3, 0.4, 0.5, 0.6, and 
0.7 $R_{\odot}$. For the calculations summarized in Figure 2, we
introduced a   
Gaussian perturbation with 
an amplitude of 1\% and a 
full-width at half maximum of 0.1 $R_{\odot}$.

\begin{figure}[t]
\centerline{\psfig{figure=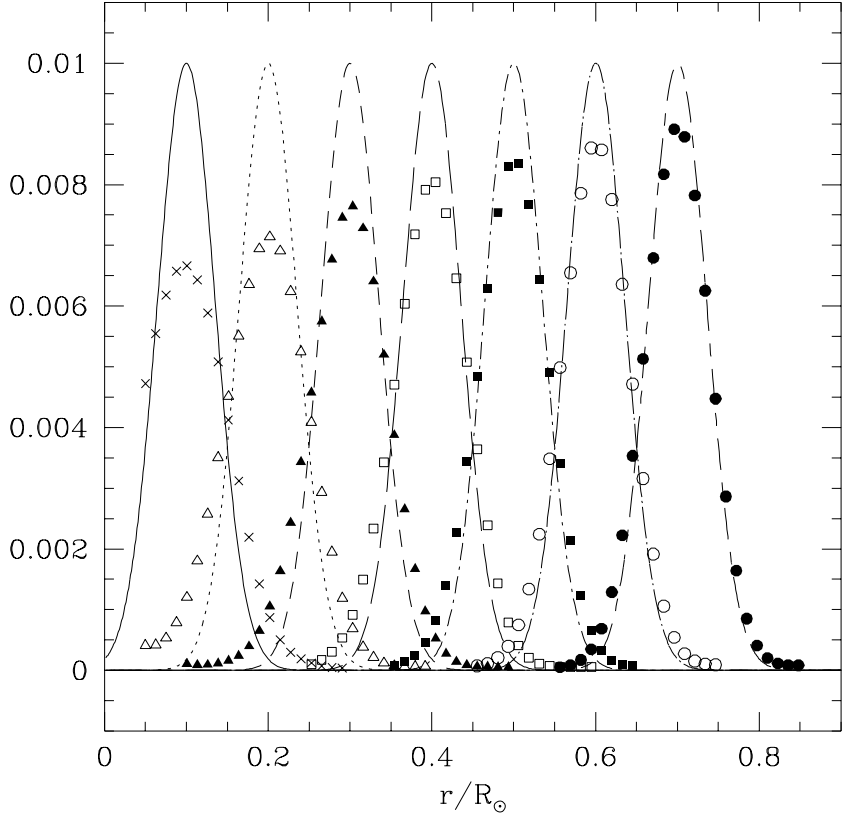,width=4.5in}}
\figcaption[bahcallfig2.ps]{Gaussian perturbation with 
$2\sigma$ widths of 0.1 $R_{\odot}$ and height 0.01
applied at various radii (shown by the lines)  and the result obtained by 
convolving the actual perturbations with the averaging kernels (shown
by the points). The perturbations and the convolved results for the
different radii are shown with different line and point styles for 
clarity.
}
\end{figure}

The effects of the perturbations are  more  easily detected in the outer
region of the sun than in the core, because relatively few modes 
penetrate the solar core and acoustic waves spend most of the travel time
in the outer layers of the Sun. 
Nevertheless, for the solar core defined as $0.1 R_{\odot}~<~r < ~0.3  
R_{\odot}$, the 1\%  peak perturbations introduced in the solar model are
detected by helioseismic inversion with an amplitude of 
about 0.7\%  after averaging over the 
inversion kernels. In the intermediate
regions,  $0.4 R_{\odot} ~<~r ~< ~0.7 R_{\odot}$,  
the 1\% peak perturbations
translate into observed discrepancies of about $0.8\%$.  

Since the standard solar model predicts sound speeds in the solar core 
that agree with the helioseismically inferred sound  speeds to better than
 $0.2\%$ rms 
(Bahcall et al. 1997), it is clear that  
any perturbation with an amplitude bigger than $(0.2/0.7) \times 1\% =  
0.3\%$ would be detectable.  Thus, the existing data limit localized  
departures from the standard model to be less than $0.3\%$ in peak 
amplitude if they extend over about $0.1 R_{\odot}$. 
The agreement 
between the standard solar model and helioseismic sound speed is 
even better, $0. 1\%$ rms, in the intermediate region. Thus we can  
place upper limits of order $(0.1/0.8) \times 1\% ~=~0.12\%$
for perturbations of width $0.1 R_{\odot}$ in the intermediate regions 
of the sun. We have verified that this conjecture is in fact correct 
to a high degree of approximation for perturbations introduced
at the 7 representative peak positions considered in Figure~2

A localized 1\% perturbation in the density in the solar core 
is recovered by the helioseismic inversion with an amplitude
of about 0.7\%. The errors in the inverted density differences 
due to uncertainties in the frequency measurements are
 about 0.22\% in the solar core and  0.15\% elsewhere.

If nothing happens to solar neutrinos after they are created in the
core of the sun, then
the existing four solar neutrino experiments suggest that the 
flux of $^7$Be neutrinos reaching the earth may be much less than is
predicted by the standard solar model.  This inference is very
difficult to understand within the constraints of standard electroweak
theory since the flux of $^7$Be neutrinos can be calculated much more
accurately than the flux of the rare $^8$B neutrinos that are observed
directly in the Kamiokande neutrino experiment (see, e.g., Bahcall
1994).  The $^7$Be and the $^8$B neutrinos are both 
produced by capture reactions on the ambient $^7$Be ions (via an
accurately known electron capture reaction 
for $^7$Be neutrinos and by a much less frequent 
and more poorly known proton capture reaction for
$^8$B neutrinos).

We have therefore tested for the sensitivity to 
ad hoc perturbations that are 
fine-tuned to give the maximum effect on the crucial $^7$Be neutrinos. 
The production of the $^7$Be neutrinos is peaked at a radius of 
$0.06R_\odot$ and FWHM of $\approx 0.075R_\odot$ (Bahcall 1989). 
We find that for a perturbation of amplitude $x$ in the sound-speed 
difference, the perceived change in the sound-speed difference is 
about $x/2$. 
Thus we can rule out  differences between
the sun and the standard solar model centered on
the $^7$Be neutrino production radius and of width $0.075R_\odot$,
which  produce sound-speed differences with 
an amplitude greater than about 0.4\%; this corresponds
to a change in the $^7$Be neutrino flux of a few percent.

\begin{figure}[t]
\centerline{\psfig{figure=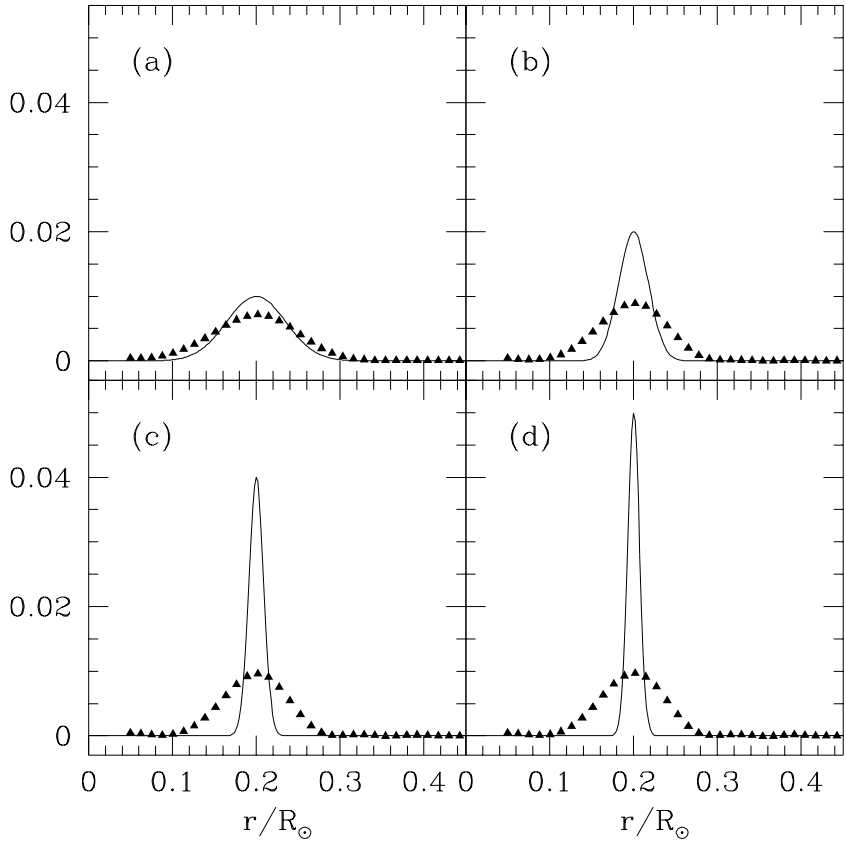,width=4.5in}}
\figcaption[bahcallfig3.ps]{Gaussian perturbations with identical areas 
applied at 0.2 $R_\odot$
(lines) and the result of convolving the perturbation with the averaging
kernels (points).  The detectability of an hypothesized perturbation 
depends primarily upon the assumed area of the perturbation. }
\end{figure}

How do the limits on  the allowable  
perturbations depend upon the assumed  
width of the  perturbations? Figure~3 illustrates the effect of varying 
the width of the perturbations while keeping fixed the area enclosed in
the $\delta c/c$ vs. $r$ plane. The figure  shows that the detectability 
of an hypothesized perturbation depends primarily on the area.
In constructing Figure~3, we have considered 
4 different perturbations all 
normalized to the same area. The peak amplitudes for the discrepancies 
introduced were, respectively, $1$\%, $2\%$, $4\%$,  and $5\%$; 
the $2\sigma$ widths were, respectively, $0.1$, $0.05$, $0.025$, 
and $0.02 R_{\odot}$.  
The narrow perturbations are considerably
broadened in the process of inversion and have a width similar to the
kernel width, and the inferred perturbation 
amplitude is correspondingly smaller.
The narrowest perturbations result in an implied discrepancy 
of almost $1\%$ while the broadest perturbation, of the same 
underlying area, gives rise to a 
discrepancy of about $0.65\%$ (see fig. 3).

\section{Summary and Discussion}

We have shown that there can be no large, reasonably 
localized differences 
between the predictions of the standard solar model and 
the sound speeds obtained from helioseismology in the region 
between $0.1 R_{\odot}$ and $0.7 R_{\odot}$.  
Numerically, we find that differences in the sound speed from the 
standard solar model  with characteristic  radial widths of  order 
$0.1 R_{\odot}$  would have been detected in
the solar core if they had peak fractional 
amplitudes greater than about $0.3\%$ and would 
have been observed in the intermediate solar region if they had peak 
fractional amplitudes in excess of about $0.1\%$.
The detectability of presumed departures from the standard solar model 
predictions depends primarily upon the assumed 
area in the $\delta c/c$ vs. $r$ plane; the 
smaller the assumed width the larger the required 
peak amplitude (see Figure~3).

In order to affect significantly the most 
important solar neutrino fluxes, 
changes in the nuclear burning temperatures of order $5\%$ to $10\%$ are 
required (see, e.g., Bahcall \& Ulmer 1996 or Bahcall 1989).  
The squares of the sound speeds are approximately proportional
to the local temperature,  so  that an uncompensated 
 $5\%$ change in the central 
temperature corresponds to about a $2.5\%$ change in the
sound speed, which is about  an order of magnitude larger than the 
upper limits set here.  We conclude, therefore, that
the standard solar model describes the structure of the sun more 
accurately than is required to predict well the solar neutrino fluxes.  
Moreover, the arguments presented here are general and the limit 
on hypothesized departures  from the standard solar model 
(e.g., from conjectured mixing, rotational instabilities, or 
magnetic fields) is small.   

Thus, the  interior of the sun is one of the 
few examples known in astronomy in 
which the conditions in the actual 
astronomical system appear to be almost as simple
as imagined by most theorists.

\section*{Acknowledgments}

The work of JNB 
was supported in part by NSF Grant No. PHY96-13835 and the 
work of SB was supported by the Danish National Research Foundation
through the establishment of the Theoretical Astrophysics Center.

\newpage

\end{document}